\documentclass[showpacs,preprintnumbers,aps]{revtex4}
\usepackage{amsfonts}
\usepackage{amssymb}
\usepackage{amsmath}
\usepackage[all]{xy}
\begin{document}
\preprint{USM-TH-167}
\newcommand{\ba}{\begin{eqnarray}}
\newcommand{\ea}{\end{eqnarray}}
\newcommand{\be}{\begin{equation}}
\newcommand{\ee}{\end{equation}}
\newcommand{\bib}{\bibitem}
\newcommand{\ed}{\end{document}}
\newcommand{\nn}{\nonumber\\}
\newcommand{\fr}{\frac}
\newcommand{\wt}{\widetilde}

\title{Duality and confinement in D=3 models driven by condensation
of topological defects}
\author{Patricio Gaete}
\email {patricio.gaete@usm.cl} \affiliation{Departamento de
F\'{\i}sica, Universidad T\'ecnica F. Santa Mar\'{\i}a,
Valpara\'{\i}so, Chile\\ and  Departamento de F\'{\i}sica,
Universidad Tecnol\'ogica Metropolitana , Santiago, Chile}
\author{Clovis Wotzasek}
\email {clovis@if.ufrj.br}
\affiliation{Instituto de F\'\i sica, Universidade Federal do Rio de
Janeiro, Brazil\\
and  Departamento de F\'{\i}sica, USACH, Chile}
\begin{abstract}
We study the interplay of duality and confinement in Maxwell-like 
three-dimensional models induced by the condensation of
topological defects driven by quantum fluctuations. To this end we
check for the confinement phenomenon, in both sides of the
duality, using the static quantum potential as a testing ground.
Our calculations are done within the framework of the
gauge-invariant but path-dependent variables formalism which are
alternative to the Wilson loop approach. Our results show that the
interaction energy contains a linear term leading to the
confinement of static probe charges.
\end{abstract}
\pacs{11.10.Ef, 11.10.Kk}
\maketitle

\section{INTRODUCTION}

This work is aimed at studying the duality symmetry for certain 3D
models coupled to sources of different dimensions that eventually
condense due to quantum fluctuations, using the Quevedo-
Trugenberger phenomenology \cite{QT} to the Julia-Toulouse
mechanism \cite{JT}. In general studies in Field Theory, the
presence of the duality between two models is verified through
their equations of motion and the algebra of the observables.
However, since quantum fluctuations may eventually drive the
condensation of topological defects destroying the duality, we
should be able to look for a distinct property to be used as a
{\it point of proof} in the condensed phase. In this paper we
propose to use the effective potential between two static charges
as such a {\it testing ground} for the existence of duality, after
quantum fluctuations drive the condensation of topological
defects. In principle one could check for this proposal studying
the interplay between confinement and duality in an $U(1)$ gauge
theory in $D$ spacetime dimensions for Maxwell-like theories of
totally anti-symmetric tensors of arbitrary rank. However, due to
technical difficulties in the computation of the effective static
potential in arbitrary $D$ dimensions we shall restrict ourselves
to the case $D=3$.

Based on the common
knowledge coming from the continuum abelian gauge theory, the
assertion that the such a theory has a confining phase may sound
strange. In fact the existence of a phase structure for the
continuum abelian U(1) gauge theory was obtained by including the
effects due to compactness of U(1) group that dramatically change
the infrared properties of the model. These results, first found
by Polyakov \cite{Polya}, have been confirmed by many distinct
techniques basically due to the contribution of the vortices into
the partition function of the theory. The condensation of these
topological defects then lead to a structural change of the
conventional vacuum of the theory into a dual superconductor
vacuum. An interesting approach to this problem has recently been
proposed by Kondo who derived the effective potential from a
partition function that includes the contribution of all
topologically non-trivial sectors of the theory \cite{Kondo}.

In a previous paper \cite{confinement} we have approached the
problem in a phenomenological way using the Julia-Toulouse
mechanism \cite{JT}, as proposed by Quevedo and Trugenberger
\cite{QT}, that considers the condensation of topological defects.
This study was undertaken for theories of compact anti-symmetric
gauge tensors of arbitrary ranks in $D$ space-time dimensions that
appear as low-energy effective field theories of strings. More
specifically, using the Quevedo-Trugenberger phenomenology
\cite{QT} we studied the low-energy field theory of a pair of
compact massless anti-symmetric tensor fields, say $A_p$ and $B_q$
with $p+q+2=D$, coupled magnetically and electrically,
respectively, to a large set of $(q-1)$-branes, characterized by
charge $e$ and a Chern-Kernel $\Lambda_{p+1}$ \cite{HL}, that
eventually condense. It has been argued that the effective theory
that results displays the confinement property. The results of
\cite{confinement} show that the phenomenological action proposed
in \cite{QT} incorporates automatically the contribution of the
condensate of topological defects to the vacuum of the model or,
alternatively, the non-trivial topological sectors as in
\cite{Kondo}.

It goes without saying that, on the grounds of the observed
electric -magnetic duality, the same result should come through if
the dual picture were adopted, i.e., by considering the reversed
couplings with a $(p-1)$-brane of charge $g$ and Chern-kernel
$\Omega_{q+1}$. Of course this seems to be mandatory if the system
displays self- duality, e.g., if the sources have the same
dimension and the tensors are of the same rank ($p = q$), which
only occurs in even dimensions. The Julia-Toulouse mechanism for
the condensation of magnetic charges, leading to confinement of
electric charges is dual to the confinement of monopoles driven by
the condensation of electric charges. However, the general results
presented in \cite{confinement} suggest that such a duality
between the electric and magnetic views should survive also for
systems not presenting self-duality, that is, when $p \neq q$.
However a clear-cut verification of this possibility is still
missing. To explicitly check for this manifestation of the duality
phenomenon in the condensed phase using the effective static
potential is our main motivation in this paper.

The way we intend to fill up this gap is by studying specific
examples in $D=3$ since they will not display self-duality. Our
calculations are done within the framework of the gauge-invariant
but path-dependent variables formalism which are alternative to
the Wilson loop approach. To consider this simpler situation seems
to be our only possibility to check for the proposal of using the
effective static potential as a {\it point of proof} for duality
in the condensed phase since for higher dimensions such a
computation becomes very messy. There is however an extra
technical point of difficulty here.  In this dimension one side of
duality involves a scalar field which is not a gauge model. In
fact it is not even a constrained theory. This poses some
difficulties applying the above formalism to compute the effective
static potential. An extension of the above mentioned method will
be presented here that solves this problem. This is a new,
although minor contribution of this work.

In the next section we discuss the duality both in the dilute and
in the condensed phases. In particular we review the main points
of the mechanism presented in \cite{confinement} using the
Quevedo-Trugenberger formalism. The phenomenological approach to
the condensation of topological defects developed in \cite{QT} and
the confinement potential for the effective theory of
\cite{confinement} are quickly reviewed in Section 2. In Section 3
we perform explicit computation for the effective potential for
the examples mentioned above. An alternative derivation of these
results together with notation and technical details in the
computation of the effective potential are included in
Appendix-\ref{VEVH} . A summary of results and future perspective
are the subject of our final section.

\section{Duality and Confinement}

\subsection{Duality in the Dilute Phase}

Consider for a moment a dual pair of massless antisymmetric
tensors, $A_p$ and $\tilde A_q$, that represent the potentials for
a pair of Maxwell-like theories, coupled to closed charged branes
of dimension $(p-1)$ and $(q-1)$.  In this dilute phase, duality
is manifest in the following sense. From the point of view of the
$A_p$ tensor, the $(p-1)$-brane is electric while the $(q-1)$-brane
is magnetic. This situation, is illustrated by the following
diagram
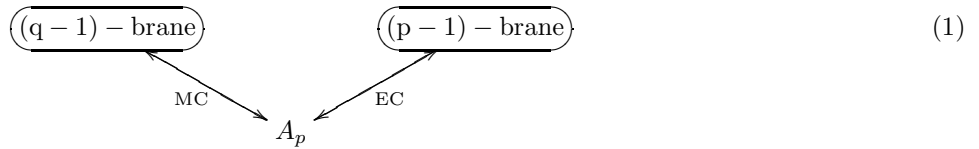
\be
\label{diagrama3}
\xymatrix{*+[F-:<10pt>]{\rm{(q-1)-brane}}  \ar[dr]& & *+[F-:<10pt>]
{\rm{(p-1)-brane}}  \ar[dl]\\
& A_p  \ar[ul]^-{\rm{MC}} \ar[ur]_-{\rm{EC}}
         }
\ee and is formally described by the following action \ba
\label{R25} {\cal S}_A &=&  \int   \frac 12 \frac{(-1)^p}{(p+1)!}
\left[F_{p+1}(A_p) - g \Lambda_{p+1}\right]^2 + e\, A_p
J^{p}(\Omega) \, . \ea where $\Lambda_{p+1}$ and $\Omega_{q+1}$
are the Chern-Kernel of the $(q-1)$-brane and the $(p-1)$-brane,
respectively. For the $D=3$ case where $p=1$ and $q=0$,
Eq.(\ref{R25}) represents the action of a vector field coupled
minimally to an electric charge and non-minimally to a magnetic
instanton. This system displays electric and magnetic symmetries
\ba \delta_E A_\mu &=& \partial_\mu \xi \;\; ; \;\; \delta_E
\Omega_\mu = \partial_\mu \xi\;\; ;\;\;\rm{electric}\nn \delta_M
A_\mu &=& g\, \chi_\mu\;\; ; \;\; \delta_M \Lambda_{\mu\nu}=
\partial_{[\mu} \chi_{\nu]} \;\; ; \;\; \rm{magnetic} \ea with
$\xi$ and $\chi_\mu$ being the corresponding gauge functions. The
new magnetic symmetry \cite{ezawa, kleinert} appears due to the
presence of magnetic sources and is manifest as the invariance of
the ``generalized" field tensor $H_{p+1} = \left[F_{p+1}(A_p) - g
\Lambda_{p+1}\right]$. The conventional (electric) gauge symmetry,
manifest by $F_{\mu\nu}(A)$ keeps the minimal coupling with the
electric pole invariant thanks to the current conservation
condition coming from the fact that we are considering closed
branes. This second term is also invariant under the magnetic
symmetry if the charges of the branes satisfy the Dirac
quantization condition, \be \label{dirac} e\, g = 2\, \pi \, n
\;\;\; ; \;\;\; n\in Z\, . \ee From the point of view of the dual
field $\tilde A_q$, the electric and magnetic couplings with the
branes have reversed character, as depicted by the diagram,
\be
\label{diagrama4}
\xymatrix{&  \tilde A_q \ar[dl]_-{\rm{EC}}  \ar[dr]^-{\rm{MC}}\\
          *+[F-:<10pt>]{\rm{(q-1)-brane}} \ar[ur] & & *+[F-:<10pt>]
          {\rm{(p-1)-brane}} \ar[ul]
         }
\ee whose action reads, \ba \label{R35} {\cal S}_{\tilde A} &=&
\int   \frac 12 \frac{(-1)^q}{(q+1)!}  \left[\tilde F_{q+1}(\tilde
A_q) - e\, \Omega_{q+1}\right]^2 + g\, \tilde A_q J^{q}(\Lambda)
\, . \ea This dual action is invariant under a corresponding pair
of electric and magnetic gauge transformation.

It is known that upon duality transformation from one picture to
the other, a brane-brane term like, \be {\cal S}_{brane-brane} =
e\; g \, \int \Lambda_{p+1} \,\epsilon^{p+1,q+1} \,\Omega_{q+1}
\ee will also be induced. Such a term however disappear from the
partition function thanks to the charge quantization condition
(\ref{dirac}) and the fact that the Chern-kernels are integer
functions, rendering the above integral as an integer that
represents the intersection number of the two branes. It is worth
noticing at this juncture that such a duality equivalence
(${*}_0$), illustrated as, \be \label{diagrama5}
\xymatrix{&  \tilde A_q \ar[dl]_-{\rm{EC}} \ar[dd] \ar[dr]^-{\rm{MC}}\\
          *+[F-:<10pt>]{\rm{(q-1)-brane}} \ar[ur] \ar[dr]& & *+[F-:<10pt>]
          {\rm{(p-1)-brane}} \ar[ul] \ar[dl]\\
         & A_p  \ar[ul]^-{\rm{MC}} \ar[ur]_-{\rm{EC}} \ar[uu]^{\mbox{$*$}_0}
         }
\ee
{\it is not} the conventional textbook duality since the potential
tensors involved are not, in general, of the same rank and,
correspondingly, the branes are not of the same dimensions. For the
particular $D=3$ case one tensor is of rank-1 (vector) and the other
is of rank-0 (scalar) while the sources are a monopole and an instanton.
The classification of these topological defects as electric or magnetic
will depend on which point of view is taken, as discussed above.
When both objects are of the same dimension (and the tensors are of the
same rank) we have self-duality. This is the usual electric-magnetic
duality discussed in the Maxwell theory in D=4, for instance, being
straightforward then that the condensation of either the electric and
the magnetic sources leads to identical physical results.

However, when the topological defects are of different dimensions
duality has to be understood in the more general sense discussed
above. Therefore without self-duality it seems doubtful that the
eventual condensation of either branes lead to the very same
physical picture. In fact, as we will show clearly in section III
by the computation of the effective static potential, the
phenomenon of confinement occurs irrespective of which brane
condense.  In particular, we will show that for $D=3$ two electric
instantons are confined when immersed in the condensate of
magnetic instantons while the electric charges experience the same
result when immersed in the condensate of magnetic monopoles.

\subsection{Duality in the Condensed Phase}

In this subsection we use the Quevedo and Trugenberger
phenomenology to study the
interplay between duality and confinement. To this end we consider
the low-energy field theory of a pair of independent massless
anti-symmetric tensor fields, say $A_p$ and $B_q$ with $p+q+2=D$,
coupled electrically (magnetically) and magnetically
(electrically) to a large set of $(q-1)$-branes, characterized by
charge $e$ and a Chern-Kernel $\Lambda_{p+1}$
(resp.,$(p-1)$-branes with charge $g$ and Chern-kernel
$\Omega_{q+1}$) \cite{HL}, that eventually condense. It will be
explicitly shown, for the D=3 example of vector and scalar fields
coupled to instantons and monopoles, that the induced effective
theories display the confinement property by providing a clear-cut
derivation of the effective potential for a pair of static, very
massive point probes immersed in the condensate.

Firstly consider that the fields $B_q$ and $A_p$ are
electrically (EC) and magnetically (MC) coupled to a $e$-charged,
$(q-1)$-brane, respectively, according to the following diagram

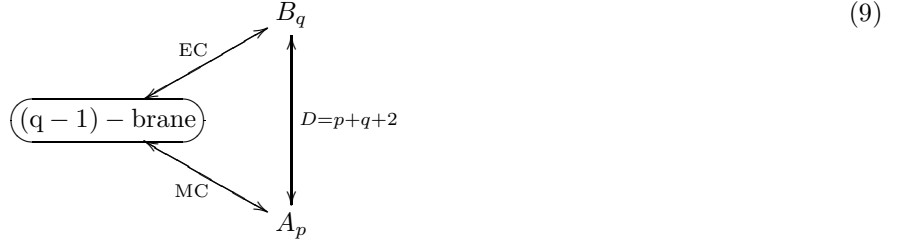
\be
\label{diagrama1}
\xymatrix{& B_q  \ar[dl]_-{\rm{EC}}\ar[dd]\\
           *+[F-:<10pt>]{\rm{(q-1)-brane}} \ar[ur] \ar[dr]& \\
        & A_p \ar[ul]^-{\rm{MC}} \ar[uu]_{D=p+q+2}
         }
\ee
The action corresponding to this situation is given by
\ba \label{R10}
{\cal S}_e &=&  \int \frac 12 \frac{(-1)^q}{(q+1)!}
\left[H_{q+1}(B_q) \right]^2 + e\, B_q J^{q}(\Lambda) +  \frac 12
\frac{(-1)^p}{(p+1)!}  \left[F_{p+1}(A_p) - e
\Lambda_{p+1}\right]^2 \, .\ea
Upon the condensation
phenomenon, the Chern-Kernel $\Lambda_{p+1}$ will become the new
massive mode of the effective theory.

Next, we consider the dual picture where the $B_q$ and $A_p$
fields are magnetically and electrically coupled to a $g$-charged,
$(p-1)$-brane, respectively, according to the following diagram
\be \label{diagrama2}
\xymatrix{ B_q  \ar[dr]^-{\rm{MC}}\ar[dd]\\
          & *+[F-:<10pt>]{\rm{(p-1)-brane}} \ar[ul] \ar[dl] \\
         A_p \ar[ur]_-{\rm{EC}} \ar[uu]^{D=p+q+2}
         }
\ee
described by the following action
\ba \label{R11}
{\cal S}_g &=&  \int \frac 12 \frac{(-1)^p}{(p+1)!}
\left[F_{p+1}(A_p) \right]^2 + g\, A_p J^{p}(\Omega) +  \frac 12
\frac{(-1)^q}{(q+1)!}  \left[H_{q+1}(B_q) - g
\Omega_{q+1}\right]^2 \ea
in which case the Chern-Kernel $\Omega_{q+1}$ will become the new
massive mode of the effective theory after the condensation.

Our compact notation here goes as follows. The field strengths are
defined as $F_{p+1}\left(A_p\right)= F_{\mu_1 \mu_2 \ldots
\mu_{p+1}}=\partial_{[\mu_1}A_{\mu_2\cdots\mu_{p+1}]}$ and
$H_{q+1} \left(B_q\right)= H_{\mu_1 \mu_2 \ldots
\mu_{q+1}}=\partial_{[\mu_1}B_{\mu_2\cdots\mu_{q+1}]}$, while the
Chern-kernels $\Lambda_{p+1}=\Lambda_{\mu_1\cdots\mu_{p+1}}$ and
$\Omega_{q+1}=\Omega_{\mu_1\cdots\mu_{q+1}}$ are totally
anti-symmetric objects of rank ($p+1$) and ($q+1$), respectively.
The conserved currents $J^q(\Lambda)$ and $J^p(\Omega)$ are given
by delta-functions over the world-volume of the ($q-1$)-brane and
($p-1$)-brane \cite{Kleinert:kx}. These conserved currents may be
rewritten in terms of the kernel $\Lambda_{p+1}$ and
$\Omega_{q+1}$ as \ba J^q(\Lambda) &=& \frac1{(p+1)!}
\epsilon^{q,\alpha ,p+1}\partial_ \alpha \Lambda_{p+1}\;
,\nonumber\\ J^p(\Omega) &=& \frac1{(q+1)!} \epsilon^{p,\alpha
,q+1}\partial_\alpha \Omega_{q+1}\; , \ea and $\epsilon^{q,\alpha
,p+1} = \epsilon^{\mu_1\ldots\mu_q,\alpha ,\nu_1\ldots\nu_{p+1}}$.

We review next how to constructed the effective interacting
action, in the condensed phase, between the anti-symmetric tensor
field $B_q$ ($A_p$) and the degrees of freedom of the condensate
$\Lambda_{p+1}$ ($\Omega_{p+1}$).
To this end, after the condensate is
integrated out, we compute the effective quantum potential for a
pair of static probe living inside the condensate, within the
framework of the gauge-invariant but path-dependent variables
formalism. This will disclose the dependence of the confinement
properties with the condensation parameters coming from the
Julia--Toulouse mechanism.

We are now ready to discuss the consequences of the Julia-Toulouse
mechanism over the action (\ref{R10}). Since the manipulations are
quite general, the result for the dual action (\ref{R11}) follows
analogously. The initial theory, before condensation, displays two
independent fields coupled to a $(q-1)$--brane. The nature of the
two couplings are however different. The $A_p$ tensor, that is
magnetically coupled to the brane, will then be absorbed by the
condensate after phase transition. On the other hand, the electric
coupling, displayed by the $B_q$ tensor, becomes a ``$B\wedge
F(\Lambda)$" topological term after condensation. Indeed, the
distinctive feature is that after condensation, the Chern-Kernel
$\Lambda_{p+1}$ is elevated to the condition of propagating field.
The new degree of freedom absorbs the degrees of freedom of the
tensor $A_p$ this way completing its longitudinal sector. The new
mode is therefore explicitly massive. Since $A_p\to \Lambda_{p+1}$
there is a change of rank with dramatic consequences. The last
term in (\ref{R10}), displaying the magnetic coupling between the
field-tensor $F_{p+1}(A_p)$ and the $(q-1)$-brane, becomes the
mass term for the new effective theory in terms of the tensor
field $\Lambda_{p+1}$ and a new dynamical term is induced by the
condensation. It is consequential that the minimal coupling of the
$B_q$ tensor becomes responsible for another contribution for the
mass, this time of topological nature. Indeed the second term
(\ref{R10}) becomes an interacting ``$BF$--term" between the
remaining propagating modes, inducing the appearance of
topological mass, in addition to the induced condensate mass.  The
final result reads \ba \label{RM10} {\cal S}_{cond} = \int
\frac{(-1)^q}{2(q+1)!} \left[H_{q+1}(B_q) \right]^2 + e\, B_q
\epsilon^{q,\alpha,p+1}\partial_\alpha \Lambda_{p+1} + \frac
{(-1)^{p+1}}{2 (p+2)!} \left[F_{p+2}(\Lambda_{p+1}) \right]^2
-\frac {(-1)^{p+1}\, (p+1)!}2 m^2 \Lambda_{p+1}^2 \ea where $m =
\Theta/ e$ where $\Theta$ represents the condensate density.

There has been a drastic change in the physical scenario. To see
this we need first to obtain an effective action for the $B_q$
tensor that includes the effects of the condensate. To this end
one integrates out the condensate field $\Lambda_{p+1}$ to obtain,
after some algebra \cite{confinement}
\begin{eqnarray}
\label{acaoeffetiva}
{\cal S}_{eff}^{(e)} =\int \frac{(-1)^{q+1}}{2\, (q+1)!} H_{q+1}(B_q)
\left(1 +\frac{e^2}{\Box + m^2} \right)H^{q+1}(B_q)\, .
\end{eqnarray}
Analogously, for the condensation of the $g$-charged brane, we obtain
from (\ref{R11}) the following effective action,
\begin{eqnarray}
\label{acaoeffetiva2}
{\cal S}_{eff}^{(g)} =\int \frac{(-1)^{p+1}}{2\, (p+1)!} F_{p+1}(A_p)
\left(1 +\frac{g^2}{\Box + \mu^2} \right)F^{p+1}(A_p)\, ,
\end{eqnarray}
where $\mu = \Theta/{g}$ is the mass parameter for this condensate.

In the next section we shall examine the duality versus
confinement issue. To this end we shall consider a specific
example involving a Maxwell field coupled electrically and
magnetically to a monopole and an instanton while the scalar
fields couples electrically and magnetically to the instanton and
the monopole.  After the condensation of the monopole we end up
with two Maxwell fields (the massive one being the condensate)
coupled topologically to each other. On the other hand after the
condensation of the instanton we end up with a massless scalar
field coupled to a massive Kalb-Ramond potential carrying the
degrees of freedom for the condensate.

\section{Interaction Energy}

Our aim in this Section is to calculate the interaction energy for
the effective theories computed above, Eqs.(\ref{acaoeffetiva})
and (\ref{acaoeffetiva2}), between appropriate external probe for
each specific model. For $D=3$ the effective actions that
naturally incorporate the contents of the dual superconductor
effects via Julia-Toulouse mechanism, for $p=1$ and $q=0$, are \ba
\label{KR151} {\cal S}_{eff}^{(pole)} = \int - \frac{1}{4}F_{\mu
\nu } \left( {1 + \frac{{e^2 }}{{\triangle^2 + m^2 }}}
\right)F^{\mu \nu }  - A_\mu J^\mu \ea and \ba {\cal
S}_{eff}^{(inst)} = \int - \frac{1}{2}\partial_\mu\phi \left( {1 +
\frac{{g^2 }}{{\triangle^2 + \mu^2 }}} \right)\partial^\mu\phi  -
\phi\, J \label{KR150} \ea where the external current will be
chosen so as to represent the presence of the two point probes.

The technique for computing the effective potential, which is
distinguished by particular attention to gauge invariance, has
been developed in \cite{Pato}.  However due to the absence of
gauge symmetry in one end of the duality, an extension of that
technique will be proposed to deal with this issue here. Other
then that our notation is defined in the Appendix-\ref{VEVH} where
the reader will also find an alternative derivation of such
results, see Eq.(\ref{ConP135}). We start with the analysis of the
theory (\ref{KR151}) that display the coupling of a vector field
to a pole. To this end consider the potential \cite{Pato},
\begin{equation}
V \equiv q\left( {{\cal A}_0 \left( {\bf 0} \right) - {\cal A}_0
\left( {\bf y} \right)} \right), \label{ConP140}
\end{equation}
where the physical scalar potential is given by
\begin{equation}
{\cal A}_0 \left( {x^0 ,{\bf x}} \right) = \int_0^1 {d\lambda }
x^i E_i \left( {\lambda {\bf x}} \right), \label{ConP145}
\end{equation}
and $i=1,2$. This follows from the vector gauge-invariant field
expression \cite{GaeteZPhys}:
\begin{equation}
{\cal A}_\mu  \left( x \right) \equiv A_\mu  \left( x \right) +
\partial _\mu\left( { - \int\limits_\xi ^x {dz^\mu  A_\mu  \left
( z \right)} } \right),
\label{ConP150}
\end{equation}
where the line integral is along a space-like path from the point
$\xi$ to $x$, on a fixed time slice, see Eq.(\ref{ConP40}). The
gauge-invariant variables (\ref{ConP150}) commute with the sole
first constraint (Gauss' law), confirming that these fields are
physical variables \cite{Dirac2}. Note that Gauss' law for the
present theory reads
\begin{equation}
\partial _i \Pi _L^i  = J^0, \label{ConP155}
\end{equation}
where $\Pi _L^i$ refers to the longitudinal part of $\Pi ^i
\equiv \left( {1 - \frac{{m^2 }}{{\nabla ^2  - e^2 }}} \right)E^i$,
and $E^i$ is the electric field. For $J^0 \left( {t,{\bf x}} \right) =
q\delta ^{\left( 2 \right)} \left( {\bf x} \right)$ the electric field is
given by
\begin{equation}
E^i  = q\left( {1 - \frac{{m^2 }}{{\nabla ^2 }}} \right)
\partial ^i G\left( {\bf x} \right), \label{ConP160}
\end{equation}
where
\be
\label{green}
G\left( {\bf x} \right) = \frac{1}{{2\pi }}K_0 \left( {M|{\bf x}|}
\right)\:\: ; \:\: M^2\equiv m^2+e^2\, ,
\ee
is the Green function for the Proca operator in $D=3$.
As a consequence, Eq.(\ref{ConP145})
becomes
\begin{equation}
{\cal A}_0 \left( {t,{\bf x}} \right) = q\left( {1 - \frac{{m^2 }}
{{\nabla ^2 }}} \right)G\left( {\bf x} \right), \label{ConP165}
\end{equation}
after subtraction of self-energy terms. Our next task is the
computation of the second term on the right-hand side of Eq.
(\ref{ConP165}). We will make use of the Green function, defined
in Eq.(\ref{ConP95}). Using this
in (\ref{ConP165}) we then obtain
\begin{equation}
\frac{G}{{\nabla ^2 }} =  - \frac{1}{{8\pi M^2 }}\left( {I_1
+ I_2 } \right),
\label{conP170}
\end{equation}
where the $I_1$ and $I_2$ terms are given by
\begin{equation}
I_1  = \int\limits_{ - \infty }^\infty  {dt} \frac{1}{{\sqrt
{1 + t^2 } }}\frac{1}{{t^2 }}\frac{{\exp \left( {iMrt} \right)}}
{{1 - {\raise0.7ex\hbox{$i$} \!\mathord{\left/
{\vphantom {i {Mrt}}}\right.\kern-\nulldelimiterspace}
\!\lower0.7ex\hbox{${Mrt}$}}}},\label{conP175}
\end{equation}
and
\begin{equation}
I_1  = \int\limits_{ - \infty }^\infty  {dt} \frac{1}{{\sqrt {1 +
t^2 } }}\frac{1}{{t^2 }}\frac{{\exp \left( {-iMrt} \right)}}{{1 +
{\raise0.7ex\hbox{$i$} \!\mathord{\left/
{\vphantom {i {Mrt}}}\right.\kern-\nulldelimiterspace}
\!\lower0.7ex\hbox{${Mrt}$}}}},\label{conP180}
\end{equation}
here $|{\bf x}|\equiv r$. The integrals (\ref{conP175}) and
(\ref{conP180}) have been explicitly computed in
Appendix-\ref{VEVH}. As a consequence, Eq. (\ref{conP170}) reduces
to
\begin{equation}
\frac{G}{{\nabla ^2 }} = \frac{r}{{4M}}. \label{conP185}
\end{equation}
Finally, making use of (\ref{conP185}) in (\ref{ConP165}), the potential
for a pair of point-like opposite charges $q$ located at ${\bf 0}$ and
${\bf L}$, becomes
\begin{equation}
V \equiv q\left( {{\cal A}_0 \left( {\bf 0} \right) - {\cal A}_0 \left(
{\bf L} \right)} \right) =  - \frac{{q^2 }}{{2\pi }}K_0 \left( {ML}
\right) + \frac{{q^2 m^2 }}{{4M}}L, \label{conP190}
\end{equation}
where $|{\bf L}|\equiv L$.
This potential displays the conventional screening part, encoded in the
Bessel function, and the linear confining potential. As expected, the
confinement disappears in the dilute phase ($m\to 0$).

Next we perform the analysis of the theory (\ref{KR150}) that
display the coupling of a scalar field to an instanton. From
(\ref{KR150}) we obtain the following equation of motion \be
\label{EM1} \left(\frac{\Box + g^2 + \mu^2}{\Box + \mu^2}\right)
\Box \phi = J\; . \ee Next, we restrict ourselves to static scalar
fields which allowed us to replace $\Box \phi = - \nabla^2 \,
\phi$. It implies that (\ref{EM1}) becomes, \be \frac{\nabla^2 -
{\cal M}^2}{\nabla^2 - \mu^2}\, \phi = - \frac{J}{\nabla^2} \ee
where now ${\cal M}^2 = \mu^2 + g^2$. From this we see that \be
\phi  = \left( {1 - \frac{{\mu^2 }}{{\nabla ^2 }}} \right)\left( {
- \frac{{J }}{{\nabla ^2  - {\cal M}^2 }}} \right) \, . \ee As
before we note that for $J \left( {t,{\bf x}} \right) = q\delta
^{\left( 2 \right)} \left( {\bf x} \right)$, the scalar field is
given by \be\label{EM5} \phi = q \left(1 -
\frac{\mu^2}{\nabla^2}\right) G\left({\bf x}\right) \ee where
$G\left({\bf x}\right)$ is the massive Green function defined
before, Eq.(\ref{green}), with $M\to {\cal M}$. It must now be
observed that Eq.(\ref{EM5}) is identical to Eq.(\ref{ConP165}),
leading to the same effective potential also for the scalar field
case, confirming that the claimed duality between the two models
persists after the condensation of the topological defects.

\section{Final Remarks}

We have used the confinement as a criterium to study duality for a
pair of antisymmetric tensors coupled to topological defects that
eventually condense. To this end we have computed the effective
static potential for a general effective theory in the condensed
phase using the Quevedo- Trugenberger formalism.  This result was
successfully used as a testing grounding for duality using the
interaction energy between two point-like probes but, for
technical reasons, we had to settle for the specific case of $D=3$
only. According to the Quevedo-Trugenberger phenomenology, the
condensation mechanism for a couple of massless antisymmetric
tensors is responsible for the appearance of mass and the jump of
rank in the magnetic sector while the electric sector becomes a BF
coupling with the condensate. The condensate absorbs and replaces
one of the tensors and becomes the new massive propagating mode
but does not couple directly to the probe charges.  The effects of
the condensation are however felt through the BF coupling with the
remaining massless tensor.  It is therefore not surprising that
they become manifest in the interaction energy for the effective
theory.   Our results show that in both sides of the duality the
interaction energy in fact contains a linear confining term. This
is an important result showing that the effective potential is a
key tool to corroborate the existence of duality, which, otherwise
are only suggested by other, very formal approaches. Extension of
this approach to check for duality in higher dimensions using the
confinement criterium is presently under investigation by the
authors.

\section{ACKNOWLEDGMENTS}

One of us (CW) would like to thank the Physics Department of the
Universidad T\'{e}cnica F. Santa Mar\'{\i}a for the invitation and
hospitality during the earlier stages of this work. This work was
supported in part by Fondecyt (Chile) Grant 1050546 (P.G.) and by
CNPq/Pronex and CAPES/Procad (Brazil).

\appendix

\section{Computation of the Confining Potential}
\label{VEVH}

Our aim in this Appendix is to recover the confining potential for
the effective theory (\ref{KR151}) computed between external probe
sources. To do this, we will compute the expectation value of the
energy operator $H$ in the physical state $\left| \Phi  \right
\rangle$ describing the sources, which we will denote by $
\left\langle H \right\rangle _\Phi$. Our starting point is the
effective Lagrangian Eq. (\ref{KR151}):
\begin{equation}
{\cal L}_{eff}  =  - \frac{1}{4}F_{\mu \nu } \left(
{1 + \frac{{m^2 }}{{\Box  + e^2 }}} \right)F^{\mu \nu }
- A_0 J^0, \label{ConP10}
\end{equation}
where $J^0$ is an external current.

Once this is done, the canonical quantization of this theory from
the Hamiltonian point of view follows straightforwardly. The canonical
momenta read $\Pi ^\mu   = - \left( {1 + \frac{{m^2 }}{{\Box
+ e^2 }}} \right)F^{0\mu }$ with the only non-vanishing canonical
Poisson brackets being
\begin{equation}
\left\{ {A_\mu  \left( {t,x} \right),\Pi ^\nu  \left( {t,y}
\right)} \right\} = \delta _\mu ^\nu  \delta \left( {x - y}
\right). \label{ConP20}
\end{equation}
Since $\Pi_0$ vanishes we have the usual primary constraint
$\Pi_0=0$, and $\Pi ^i  = \left( {1 + \frac{{m^2 }}{{\Box +
e^2 }}} \right)F^{i0}$. The canonical Hamiltonian is thus
\begin{equation}
H_C  = \int {d^3 } x\left\{ { - \frac{1}{2}\Pi ^i \left( {1 +
\frac{{m^2 }}{{\Box  + e^2 }}} \right)^{ - 1} \Pi _i  + \Pi
^i \partial _i A_0  + \frac{1}{4}F_{ij} \left( {1 + \frac{{m^2
}}{{\Box  + e^2 }}} \right)F^{ij}  + A_0 J^0 }
\right\}.\label{ConP25}
\end{equation}
Time conservation of the primary constraint $ \Pi _0$ leads to the
secondary Gauss-law constraint
\begin{equation}
\Gamma _1 \left( x \right) \equiv \partial _i \Pi ^i - J^0 =
0.\label{ConP30}
\end{equation}
The preservation of $\Gamma_1$ for all times does not give rise to
any further constraints. The theory is thus seen to possess only
two constraints, which are first class, therefore the theory
described by $(\ref{ConP10})$ is a gauge-invariant one. The extended
Hamiltonian that generates translations in time then reads $H =
H_C  + \int {d^2 } x\left( {c_0 \left( x \right)\Pi _0 \left( x
\right) + c_1 \left( x \right)\Gamma _1 \left( x \right)}
\right)$, where $c_0 \left( x \right)$ and $c_1 \left( x \right)$
are the Lagrange multiplier fields. Moreover, it is
straightforward to see that $\dot{A}_0 \left( x \right)= \left[
{A_0 \left( x \right),H} \right] = c_0 \left( x \right)$, which is
an arbitrary function. Since $ \Pi^0 = 0$ always, neither $ A^0 $
nor $ \Pi^0 $ are of interest in describing the system and may be
discarded from the theory. Then, the Hamiltonian takes the form
\begin{equation}
H = \int {d^2 x\left\{ { - \frac{1}{2}\Pi _i \left( {1 +
\frac{{m^2 }}{{\Box  + e^2 }}} \right)^{ - 1} \Pi ^i  +
\frac{1}{4}F_{ij} \left( {1 + \frac{{m^2 }}{{\Box  + e^2 }}}
\right)F^{ij}  + c\left( x \right)\left( {\partial _i \Pi ^i  -
J^0 } \right)} \right\}}, \label{ConP35}
\end{equation}
where $c(x) = c_1 (x) - A_0 (x)$.

The quantization of the theory requires the removal of nonphysical
variables, which is done by imposing a gauge condition such that
the full set of constraints becomes second class. A convenient
choice is found to be \cite{Pato}
\begin{equation}
\Gamma _2 \left( x \right) \equiv \int\limits_{C_{\xi x} } {dz^\nu
} A_\nu \left( z \right) \equiv \int\limits_0^1 {d\lambda x^i }
A_i \left( {\lambda x} \right) = 0, \label{ConP40}
\end{equation}
where  $\lambda$ $(0\leq \lambda\leq1)$ is the parameter
describing the spacelike straight path $ x^i = \xi ^i  + \lambda
\left( {x - \xi } \right)^i $, and $ \xi $ is a fixed point
(reference point). There is no essential loss of generality if we
restrict our considerations to $ \xi ^i=0 $. In this case, the
only non-vanishing equal-time Dirac bracket is
\begin{equation}
\left\{ {A_i \left( x \right),\Pi ^j \left( y \right)} \right\}^ *
=\delta{ _i^j} \delta ^{\left( 2 \right)} \left( {x - y} \right) -
\partial _i^x \int\limits_0^1 {d\lambda x^j } \delta ^{\left( 2
\right)} \left( {\lambda x - y} \right). \label{ConP45}
\end{equation}
In passing we recall that the transition to quantum theory is made
by the replacement of the Dirac brackets by the operator
commutation relations according to
\begin{equation}
\left\{ {A,B} \right\}^ *   \to \left( { - i} \right)\left[ {A,B}
\right]. \label{ConP50}
\end{equation}

We now turn to the problem of obtaining the interaction energy
between point-like sources in the model under consideration.
The state $\left| \Phi  \right\rangle$ representing the sources is
obtained by operating over the vacuum with creation/annihilation
operators. We want to stress that, by construction, such states are
gauge invariant. In the case at hand we consider the gauge-invariant
stringy $\left|{\overline \Psi  \left( \bf y \right)\Psi \left
( {\bf y^ \prime }\right)} \right\rangle$, where a fermion is
localized at ${\bf y}\prime$ and an anti-fermion at ${\bf y}$
as follows \cite{Dirac2},
\begin{equation}
\left| \Phi  \right\rangle  \equiv \left| {\overline \Psi  \left(
\bf y \right)\Psi \left( {\bf y}\prime \right)} \right\rangle  =
\overline \psi \left( \bf y \right)\exp \left(
{iq\int\limits_{{\bf y}\prime}^{\bf y} {dz^i } A_i \left( z
\right)} \right)\psi \left({\bf y}\prime \right)\left| 0
\right\rangle, \label{ConP60}
\end{equation}
where $\left| 0 \right\rangle$ is the physical vacuum state and
the line integral appearing in the above expression is along a
space-like path starting at ${\bf y}\prime$ and ending $\bf y$, on
a fixed time slice. It is worth noting here that the strings
between fermions have been introduced in order to have a
gauge-invariant function $\left| \Phi  \right\rangle $. In other
terms, each of these states represents a fermion-antifermion pair
surrounded by a cloud of gauge fields sufficient to maintain gauge
invariance. As we have already indicated, the fermions are taken
to be infinitely massive (static).

From our above discussion, we see that $\left\langle H
\right\rangle _\Phi$ reads
\begin{equation}
\left\langle H \right\rangle _\Phi   = \left\langle \Phi
\right|\int {d^2 x\left\{ { - \frac{1}{2}\Pi _i \left( {1 +
\frac{{m^2 }}{{\Box  + e^2 }}} \right)^{ - 1} \Pi ^i  +
\frac{1}{4}F_{ij} \left( {1 + \frac{{m^2 }}{{\Box  + e^2 }}}
\right)F^{ij} } \right\}}\left| \Phi  \right\rangle. \label{ConP55}
\end{equation}
Consequently, we can write
Eq.(\ref{ConP55}) as
\begin{equation}
\left\langle H \right\rangle _\Phi   = \left\langle \Phi
\right|\int {d^2 x } \left\{ { - \frac{1}{2}\Pi _i \left( {1 -
\frac{{m^2 }}{{\nabla ^2  - e^2 }}} \right)^{ - 1} \Pi ^i }
\right\}\left| \Phi  \right\rangle, \label{ConP65}
\end{equation}
where, in this static case, $\Box = - \nabla ^2$. Observe
that when $m=0$ we obtain the pure Maxwell theory, as already
mentioned.  From now on we will suppose $m\neq 0$.

Next, from our above Hamiltonian analysis, we note that
\begin{equation}
\Pi _i \left( x \right)\left| {\overline \Psi  \left( {\bf y}
\right)\Psi \left( {\bf y^\prime} \right)} \right\rangle  =
\overline \Psi \left( {\bf y} \right)\Psi \left( {\bf y^\prime}
\right)\Pi _i \left( x \right)\left| 0 \right\rangle  +
q\int\limits_{\bf y}^{\bf y^\prime} {dz_i \delta ^{(2)} \left(
{{\bf z} - {\bf x}} \right)\left| \Phi \right\rangle
}.\label{ConP70}
\end{equation}
As a consequence, Eq.(\ref{ConP65}) becomes
\begin{equation}
\left\langle H \right\rangle _\Phi   = \left\langle H
\right\rangle _0  + V^{\left( 1 \right)}  + V^{\left( 2 \right)},
\label{ConP75}
\end{equation}
where $\left\langle H \right\rangle _0  = \left\langle 0
\right|H\left| 0 \right\rangle$. The $V^{\left( 1 \right)}$ and
$V^{\left( 2 \right)}$ terms are given by:
\begin{equation}
V^{\left( 1 \right)}  =  - \frac{{q^2 }}{2}\int {d^2 x} \int_{\bf
y}^{\bf y^\prime} {dz^\prime_i } \delta ^{\left( 2 \right)} \left(
{x - z^\prime} \right)\frac{1}{{\nabla _x^2  - M^2 }}\nabla _x^2
\int_{\bf y}^{\bf y^\prime} {dz^i } \delta ^{\left( 2 \right)}
\left( {x - z} \right), \label{ConP80}
\end{equation}
and
\begin{equation}
V^{\left( 2 \right)}  =   \frac{{q^2 m^2}}{2}\int {d^2 x}
\int_{\bf y}^{\bf y^\prime} {dz^\prime_i } \delta ^{\left( 2
\right)} \left( {x - z^\prime} \right)\frac{1}{{\nabla _x^2  - M^2
}} \int_{\bf y}^{\bf y^\prime} {dz^i } \delta ^{\left( 2 \right)}
\left( {x - z} \right), \label{ConP85}
\end{equation}
where $ M^2\equiv{m^2+ e^2} $ and the integrals over $z^i$ and
$z^\prime_i$ are zero except on the contour of integration.

The $V^{\left( 1 \right)}$ term may look peculiar, but it is just
the familiar Bessel interaction plus self-energy terms. In effect,
expression (\ref{ConP80}) can also be written as
\begin{equation}
V^{\left( 1 \right)}  = \frac{{q^2 }}{2}\int_{\bf y}^{{\bf
y}^{\prime}  } {dz_i^{\prime}}\partial _i^{z^{\prime}} \int_{\bf
y}^{{\bf y}^{\prime}} {dz^i }\partial _z^i G\left( {{\bf
z}^{\prime},{\bf z}} \right), \label{ConP90}
\end{equation}
where $G$ is the Green function
\begin{equation}
G({\bf z}^{\prime}  ,{\bf z}) = \frac{1}{{2\pi }}K_0 \left(
{M|{\bf z}^{\prime}  - {\bf z}  |} \right). \label{ConP95}
\end{equation}
Employing Eq.(\ref{ConP95}) and remembering that the integrals
over $z^i$ and $z_i^{\prime}$ are zero except on the contour of
integration, expression (\ref{ConP90}) reduces to the familiar
Bessel interaction after subtracting the self-energy terms, that
is,
\begin{equation}
V^{\left( 1 \right)}  = - \frac{q^2}{{2\pi }}K_0 \left( {M|{\bf y}
- {\bf y}^{\prime}  |} \right). \label{ConP100}
\end{equation}

We now turn our attention to the calculation of the $V^{\left( 2
\right)}$ term, which is given by
\begin{equation}
V^{\left( 2 \right)}  = \frac{{q^2 m^2 }}{2}\int_{\bf y}^{{\bf
y}^{\prime}  } {dz^{{\prime} i} } \int_{\bf y}^{{\bf y}^{\prime} }
{dz^i } G({\bf z}^{\prime} ,{\bf z}). \label{ConP105}
\end{equation}
It is appropriate to observe here that the above term is similar
to the one found for the system consisting of a gauge field
interacting with an external background\cite{GaeteEnero}.
Notwithstanding, in order to put our discussion into context it is
useful to summarize the relevant aspects of the calculation
described previously \cite{GaeteEnero}. In effect, as was explained in
Ref. \cite{GaeteEnero}, by using the integral representation of the
Bessel function
\begin{equation}
K_0 \left( x \right) = \int\limits_0^\infty  {\cos (x\sinh t)dt =
} \int\limits_0^\infty  {\frac{{\cos (xt)}}{{\sqrt {t^2  + 1} }}}
dt, \label{ConP110}
\end{equation}
where $x>0$, expression (\ref{ConP105}) can also be written as
\begin{equation}
V^{\left( 2 \right)}   = \frac{{q^2 m^2 }}{{2\pi M^2
}}\int\limits_0^\infty  {dt\frac{1}{{t^2 }}} \frac{1}{{\sqrt {t^2
+ 1} }}\left( {1 - \cos \left( {MLt} \right)} \right),
\label{ConP115}
\end{equation}
where $L\equiv|{\bf y}-{\bf {y^\prime}}|$.
Now let us calculate integral (\ref{ConP115}). For this purpose we
introduce a new auxiliary parameter $\varepsilon$ by making in the
denominator of integral (\ref{ConP115}) the substitution
$t^2\rightarrow t^2+\varepsilon^2$. Thus it follows that
\begin{equation}
V^{\left( 2 \right)} \equiv \lim _ {\varepsilon  \to 0}
{\widetilde V}^{\left( 2 \right)}= \lim _{\varepsilon \to
0}\frac{{q^2 m^2 }}{{2\pi M^2 }}\int\limits_0^\infty
{\frac{{dt}}{{t^2  + \varepsilon ^2 }}} \frac{1}{{\sqrt {t^2  + 1}
}}\left( {1 - \cos \left( {MLt} \right)} \right). \label{ConP120}
\end{equation}
A direct computation on the $t$-complex plane yields
\begin{equation}
{\widetilde V}^{\left( 2 \right)} = \frac{{q^2 m^2 }}{{4 M^2 }}
\left( {\frac{{1 - {\mathop{\rm e}\nolimits} ^{ - ML\varepsilon }
}}{\varepsilon }} \right)\frac{1}{{\sqrt {1 - \varepsilon ^2 } }}
. \label{ConP125}
\end{equation}
Taking the limit $\varepsilon  \to 0$, expression (\ref{ConP125})
then becomes
\begin{equation}
V^{\left( 2 \right)}  = \frac{{q^2 m^2 }}{{4M }}|{\bf y} - {{\bf
y}^\prime}|. \label{ConP130}
\end{equation}
From Eqs.(\ref{ConP100}) and (\ref{ConP130}), the corresponding
static potential for two opposite charges located at ${\bf y}$ and
${\bf y^\prime}$ may be written as
\begin{equation}
V(L) = - \frac{{q^2 }}{{2\pi }}K_0 \left( {ML} \right) +
\frac{{q^2 m^2 }}{{4M}}L, \label{ConP135}
\end{equation}
where $L\equiv|{\bf y}-{\bf {y^\prime}}|$.

\end{document}